# Ultrafast Self-powered Visible Blind UV Photodetectors based on MgZnO Vertical Schottky Junction in Crossbar Geometry

*Amit K Das\*, V. K. Sahu, R. S. Ajimsha and P. Misra*

Oxide Nano-Electronics Laboratory, Laser Material Processing Division, Raja Ramanna Centre for Advanced Technology, Indore – 452 013, India

*Corresponding author email: amitdas@rrcat.gov.in

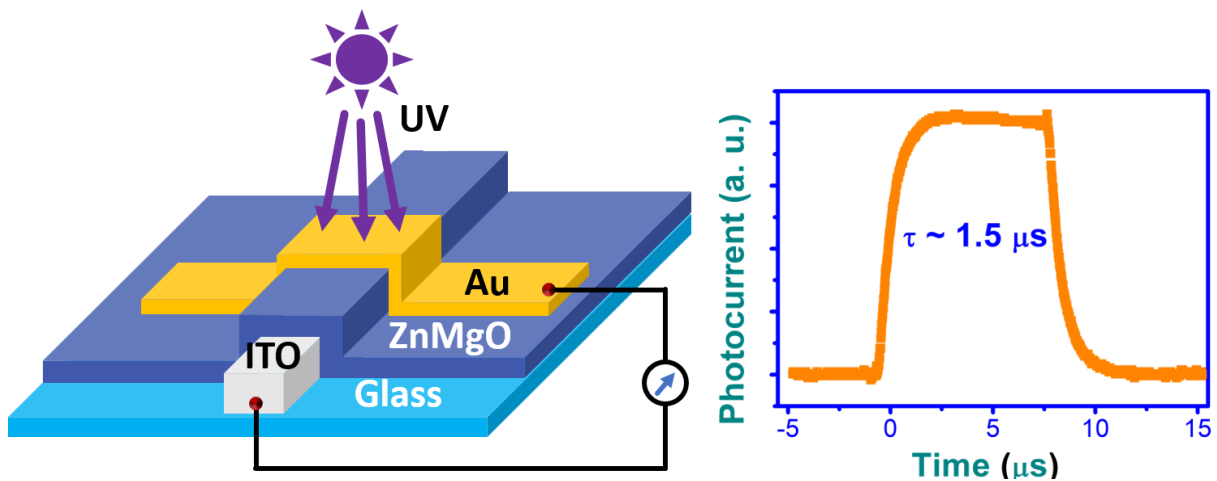

## Abstract

In order to achieve ultrafast response in MgZnO based self-powered Schottky type photodetectors, it is crucial to decrease both the junction capacitance and carrier transit time. To meet these criteria, Au/MgZnO/ITO Schottky junction photodetectors have been realised in crossbar pattern, wherein the thickness of the MgZnO thin film deposited on patterned ITO-glass substrate is ~ 200 nm and the cross-sectional area of the devices is 0.032 $mm^2$. The semi-transparent 10 nm Au electrode on top of the MgZnO film serves as the Schottky electrode through which light enters the devices. The vertical geometry of the crossbar pattern and the associated small device cross-section results in a low junction capacitance of the devices of ~ 27 pF at zero bias, which in turn produces very fast visible blind self-powered ultraviolet (UV) photoresponse with both the rise and fall times of ~ 1.5 µs. The devices also demonstrate a peak responsivity of ~ 49 mA/W at ~ 280 nm with a cut-off wavelength of 336 nm. These Au/MgZnO/ITO Schottky photodetectors in crossbar pattern, with optimized device area, could be useful in applications requiring fast response, such as UV communication and UV imaging.

**Introduction**

Fast visible blind UV photodetectors are essential for ultraviolet communications protocols which have numerous advantages such as non-line-of-sight (NLOS) and radio silent operation, anti-interference, low background noise, high bandwidth etc.[1]–[3] Apart from UV communication, fast visible blind UV detectors are also useful in a number of applications such as missile plume and stealth aircraft detection, high-capacity UV storage, detection of fire and spark in high voltage transmission lines and equipment etc.[4]–[6] Primary requirement for developing visible blind UV detectors is semiconductors that have wide bandgap so as to not respond to visible light. $Mg_xZn_{1-x}O$, which is an alloy of insulator MgO and wide bandgap semiconductor ZnO, is a suitable material in this regard as the bandgap of $Mg_xZn_{1-x}O$ can be tailored from 3.3 eV, which is the bandgap of ZnO to ~ 5.5 eV by simply varying $x$, the concentration of Mg as compared to Zn.[7], [8] Other important factors that make $Mg_xZn_{1-x}O$ suitable for photodetector applications are ease of growth of thin films, environment friendliness and radiation hardness.[9]–[11] Different types of photodetectors based on MgZnO such as p-n homojunction, p-n heterojunction, Schottky type, metal-semiconductor-metal photoconducting type etc. have been reported in literature.[12] Among these, p-n junction and Schottky type photodetectors can work in self-powered mode, meaning they can work without external power supply due to the presence of the built-in field at the junction.[10] Since it is difficult to achieve stable p-type doping in MgZnO, Schottky junction based visible blind UV detectors are promising for realizing MgZnO based self-powered visible blind UV detectors.[10], [12] The Schottky junction photodetectors have the added advantage that their response can be fast as they are majority carrier devices. Schottky junctions of Au, Ag, Pt and PEDOT:PSS on MgZnO films have been utilized to fabricate UV photodetectors with varying degree of performance.[10], [13]–[15] Among these the fastest response (rise time ~ 20 μs and decay time 300 μs) was obtained with comb shaped Au Schottky electrode on 600 nm MgZnO thin film.[10]

It is well known from literature that the response speed of Schottky junction photodetectors primarily depends on two parameters – the junction capacitance and the carrier transit time.[16] The carrier transit time can be reduced by making the device in vertical geometry together with reducing the thickness of the MgZnO film and enhancing the ratio of depletion width to the thickness. The junction capacitance can be reduced by decreasing the junction cross-section and increasing the depletion width. The depletion width depends on the concentration of electrons in MgZnO which can be lowered by depositing the film in presence of adequate oxygen so that there are fewer oxygen vacancies in the MgZnO film.[4], [17] Therefore, to obtain fast response speed in MgZnO based Schottky photodetectors, it is necessary to fabricate the devices in vertical geometry with small cross-section together with fewer oxygen vacancies in the MgZnO thin films. However, these aspects are largely ignored in the existing literatures that describe response speed of MgZnO based Schottky photodetectors. In this report, we have fabricated MgZnO based vertical Schottky diode with Au as the Schottky electrode and ITO as the ohmic electrode with the goal of achieving fast response. The devices have been fabricated in cross-bar geometry with individual device cross sectional area of ~ 0.032 mm$^2$. The devices showed both photoresponse rise and fall time of ~ 1.5 µs, which is the fastest photoresponse obtained so far in MgZnO based Schottky junction UV photodetectors. The experimental details and the results of the study are presented and discussed in detail in the following sections.

**Experimental**

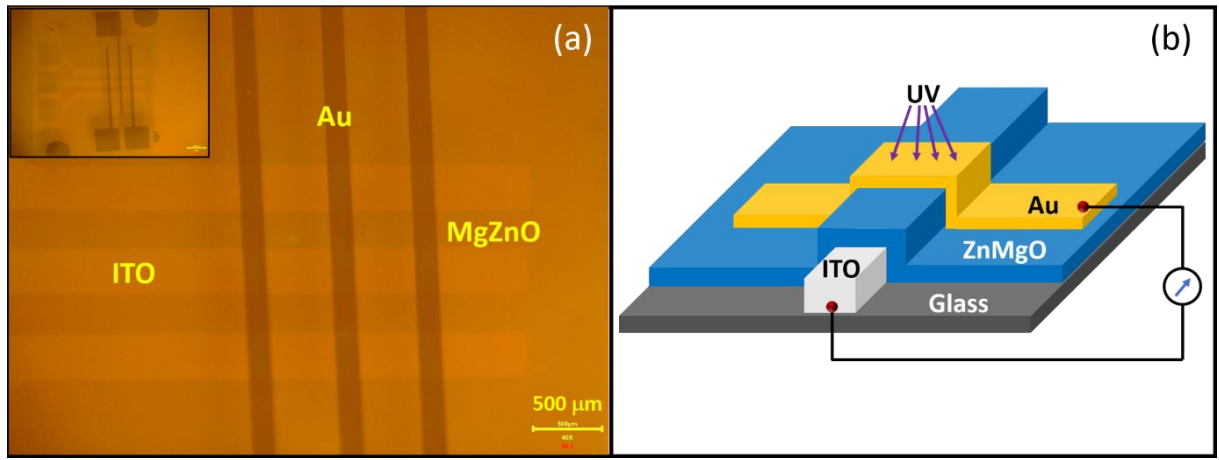

**Figure 1:** *(a) Optical microscopic image of the Au/MgZnO/ITO vertical Schottky junction photodetectors in crossbar pattern and (b) the 3D schematic of a single device and the electrical connections.*

The Au/MgZnO/ITO Schottky junction photodetectors have been fabricated in crossbar pattern as can be seen in the optical microscopic image of the devices in figure 1(a). The bottom ITO electrodes of width ~ 325 µm were patterned by photolithography followed by etching of 20 ohm/square ITO coated glass substrate in 9M HCL.[18] The ITO electrodes served as the ohmic contact to the MgZnO thin film which was subsequently deposited on the ITO electrodes by RF magnetron sputtering at 100 W power using a ceramic MgZnO target containing about 30% Mg. The sputtering process was carried out at a substrate temperature of 200$^0$C in a mixture of argon and oxygen gas at a total pressure of 4.1x10$^{-2}$ mbar, where the flow rate of Ar and O$_2$ were maintained at 21 and 15 sccm respectively. The top semi-transparent Schottky electrodes of gold of width ~ 100 μm were fabricated by photolithography followed by deposition of Au thin film by RF sputtering and lift-off. As shown in figure 1(a), total 9 devices have been fabricated on a substrate in cross bar geometry, each one of which can be selected for probing by connecting the measurement device to one horizontal (ITO) and one vertical (Au) electrode. The inset of figure 1(a) shows the complete array of 9 Au/MgZnO/ITO vertical Schottky diodes in cross-bar pattern along with the electrode pad areas. The 3D schematic of a single Au/MgZnO/ITO Schottky junction photodetector and the electrode connections are depicted in figure 1(b). A spectroscopic reflectometer (make: Angstrom-Sun, model: TFProbe) and a stylus-based thickness profiler both have been used to measure the thicknesses of the MgZnO thin film and Au electrode. The thickness of the MgZnO thin film was estimated by both the techniques to be in the range from ~ 200 - 220 nm, while that of the Au top electrode has also been estimated to be ~ 7-10 nm, so that it can let through incident light to reach the Schottky junction between Au and MgZnO. For estimation of bandgap of the MgZnO thin film and transmission percentage of the Au electrode, MgZnO and Au thin films have been deposited on Al$_2$O$_3$ substrate as well. The bandgap of the MgZnO thin film has been estimated from transmission spectra measured with a UV-visible spectrophotometer (make: LAB India, model: UV3200). The same instrument has also been used to measure the transmission spectra of the 10 nm Au thin film. For determining the composition by EDX measurements, MgZnO thin film has been deposited on Si substrate under identical condition to that in the case of ITO substrate. Surface SEM measurements have also been carried out on the MgZnO thin film deposited on ITO substrate to see the surface morphology. SEM and EDX measurements have been carried out using Carl Zeiss make SEM machine (model: Zigma). To study the crystalline quality of the MgZnO thin film deposited on ITO-glass substrate, X-ray diffraction measurements have been carried out using a Bruker make XRD machine (model: Discover

D8). The electrical characteristics of the Au/MgZnO/ITO vertical Schottky diodes have been measured by using a Kiethley make source-measure unit (model: 2450) in the measurement setup shown schematically in figure 1(b). A 70 W xenon lamp (make: Oriel) coupled to a quarter meter monochromator (make: Solar TII) has been utilized as the light source for photoresponse measurements. The junction capacitance of the devices at zero bias have been measured with a LCR meter (make: HIOKI, model: IM3536). The temporal photoresponse of the devices has been recorded by exciting the devices with a 2 W UV LED emitting at 275 nm (make: Osram Oslon) that was powered by a Tektronix make pulse generator (model: AFG3152C). The Au/MgZnO/ITO UV photodetectors were connected to an in-house built 500 kHz transimpedance amplifier with a gain of ~ $2 \times 10^5$ V/A that was in turn connected to an oscilloscope (make: Agilent, model: DSO6054A) to record the temporal photoresponse. The excitation source i.e., the UV LED was turned ON and OFF alternately at a frequency of 60 kHz to record the photoresponse of the Au/MgZnO/ITO devices.

**Result and discussion**:

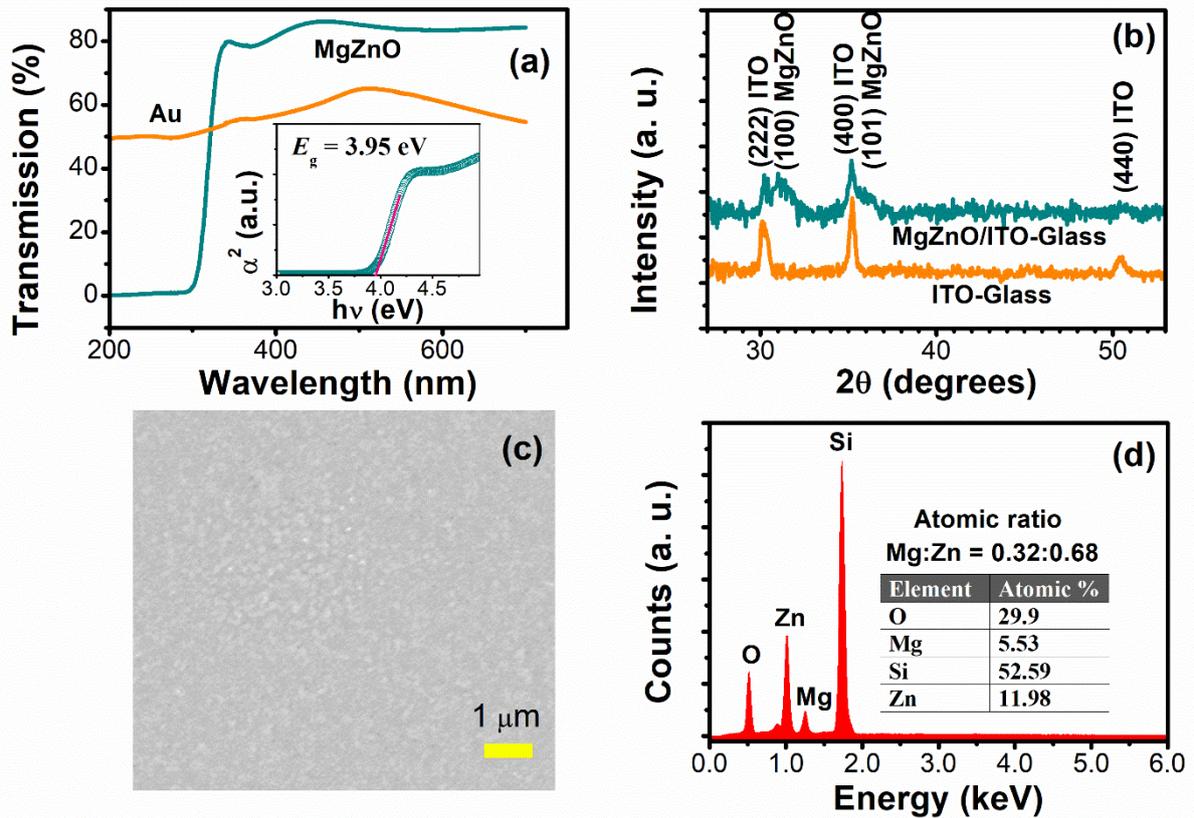

**Figure 2:** *(a) UV-Visible transmission spectra of MgZnO thin film and 9 nm Au thin film on sapphire substrates, (inset) $\alpha^2$ as a function of photon energy ($h\nu$) for the MgZnO thin film, (b) XRD patterns of the ITO-glass substrate (orange) and the MgZnO thin film on ITO-glass substrate (dark cyan), (c) SEM image of MgZnO thin film on ITO-glass substrate and (d) the EDAX spectra of the MgZnO thin film on Si.*

The transmission spectra of the MgZnO thin film deposited on sapphire substrate is shown in figure 2 (a). As can be seen, the MgZnO thin film is transparent in the visible spectral region with ~ 85% transmission. The transmission percentage drops sharply after ~ 330 nm and becomes almost negligible for wavelength less than 300 nm due to band edge absorption in MgZnO film. The inset shows $\alpha^2$ as a function of photon energy ($h\nu$) for the film, from which the bandgap of the MgZnO has been estimated to be ~ 3.95 eV. The figure 2(a) also shows the transmission spectra of the Au thin film with a thickness of ~ 7-10 nm, which is more than 50% transparent in the entire spectral range from 700 – 190 nm, implying that it can transmit significant amount of the incident UV radiation to the Au/MgZnO Schottky junction. In order to investigate the crystalline structure of the MgZnO thin films deposited on ITO-glass substrate, X-ray diffraction measurements were carried out on the devices, the results of which are depicted in figure 2(b). The same figure also shows the XRD pattern of ITO-glass substrate for the purpose of comparison. It can be seen that in the XRD spectra of the MgZnO thin film, the peaks corresponding to (100) and (101) planes of hexagonal wurtzite MgZnO are present, apart from the peaks corresponding to ITO, implying polycrystalline nature of the wurtzite MgZnO thin film deposited on ITO-glass substrate. [19]–[22] The broadness of the XRD peaks is possibly due to small crystallite size. Figure 1(c) depicts the surface SEM image of the MgZnO thin film deposited on ITO-glass substrate. The surface is uniform and relatively smooth showing the presence of MgZnO grains. The composition of the MgZnO thin film has been determined from EDX spectra of the MgZnO thin film deposited on Si substrate, shown in the figure 2(d). From the EDX spectra it has been calculated that the atomic percentage of Mg as compared to Zn in the MgZnO thin film is ~ 32 %, so that the composition of MgZnO thin film cam be written as $Mg_{0.32}Zn_{0.68}O$. The concentration of Mg is commensurate with the bandgap observed in figure 2(a), as can be inferred from the literature. [23] Moreover, it can be also seen from the inset of figure 2(d) that the atomic percent of oxygen is higher than that of Mg and Zn combined, implying oxygen rich MgZnO thin film.

The I-V characteristics of the $Au/Mg_{0.3}Zn_{0.7}O/ITO$ Schottky junction photodetectors are depicted in Figure 3(a). In dark the I-V curve (orange coloured) looks similar to that of a typical rectifying junction with a rectification ratio of ~ 60 at 1 V. The values of dark current at 0.5 and 1 V reverse bias are ~ 6 pA and 10 pA. The I-V characteristics of the device in 275 nm UV illumination at an optical power density of ~ 0.09 mW/cm$^2$ is also shown in figure 3 in dark cyan color. It is evident that the current at a reverse bias of 1 V increases by about 3 orders of magnitude in presence of the UV light. Moreover, the value of the photocurrent at zero bias is found to be ~ 200 pA, implying that the device can work in photovoltaic mode and it is possible to utilize the Au/MgZnO/ITO Schottky junction as a self-powered photodetector.[10] This photovoltaic nature can also be seen from figure 3(b), where the I-V curves in dark and in UV are presented in linear scale in zoomed-in state. It clearly demonstrates the generation of open circuit voltage ($V_{OC}$) of about 0.15 V and short circuit current ($I_{SC}$) of about 200 pA under UV illumination. It is worthwhile to note here that due to the small area of the device, the optical power incident on the devices is ~ 40 nW.

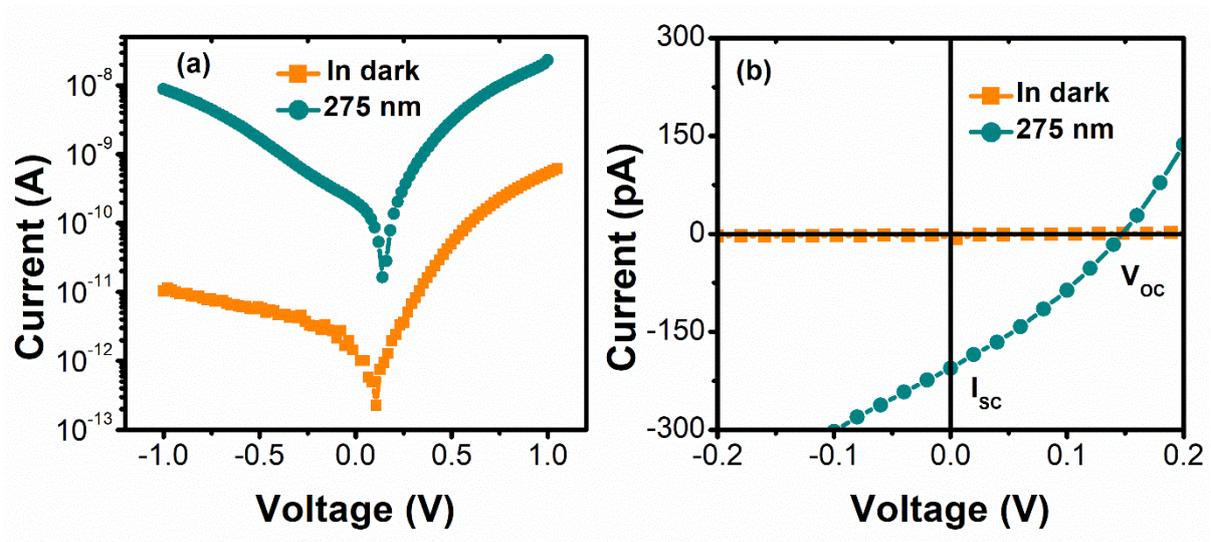

*Figure 3: (a) The current-voltage (I-V) characteristics of the Au/MgZnO/ITO Schottky junction in semi-log scale in dark (orange filled squares) and in presence of 0.09 mW/cm² of 275nm UV illumination (dark cyan filled circles). (b) The zoomed-in I-V curves in linear scale to clearly demonstrate the photovoltaic character.*

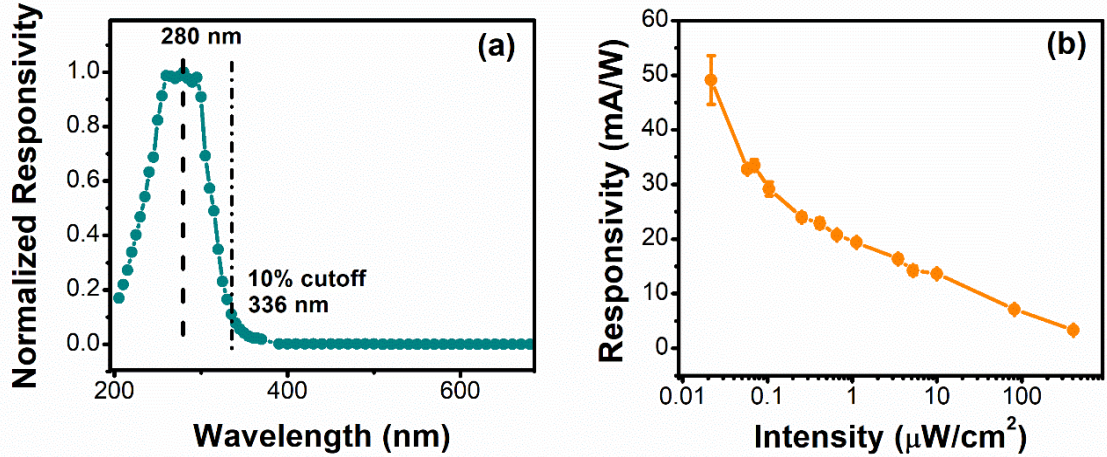

**Figure 4**: (a) Spectral responsivity of the Au/MgZnO/ITO Schottky junction photodetector in zero bias and (b) responsivity at zero bias as a function of incident light intensity measured at 275 nm.

The normalized spectral responsivity of the Au/MgZnO/ITO Schottky junction photodetectors at zero bias is shown in figure 4(a). As can be seen from the figure that the responsivity peaks in the wavelength range from 290 -260 nm with 10% cut-off at 336 nm, which matches with the transmission spectra of the MgZnO thin film depicted in figure 2(a). There is practically negligible photoresponse in the visible spectral region indicating that the device can be used as visible blind UV photodetector. Figure 4(b) shows the responsivity at zero bias as a function of the incident intensity of 275 nm UV light. As can be seen from the figure that with decreasing UV intensity from 410 µW/cm² to ~ 22 nW/cm², the responsivity at 275 nm increases monotonically from ~ (3.3 ± 0.2) mA/W to ~ (49 ± 4) mA/W. The peak responsivity of ~ 49 mA/W at zero bias is quite high and comparable to that reported in literature.[10] The decrease in responsivity of the Au/MgZnO/ITO devices with increasing intensity of UV light is due to generation of large density of photogenerated carriers at higher incident light intensity that increases the recombination probability, thereby reducing the photoresponsivity.[10] It is also worthwhile to note here that the device can measure very low power density of ~ 22 nW/cm² so that it can useful for detection of weak UV signal. UV to visible rejection ratio of the device defined as the ratio of responsivities at 280 nm to that at 400 nm is ~ $3.1 \times 10^3$.

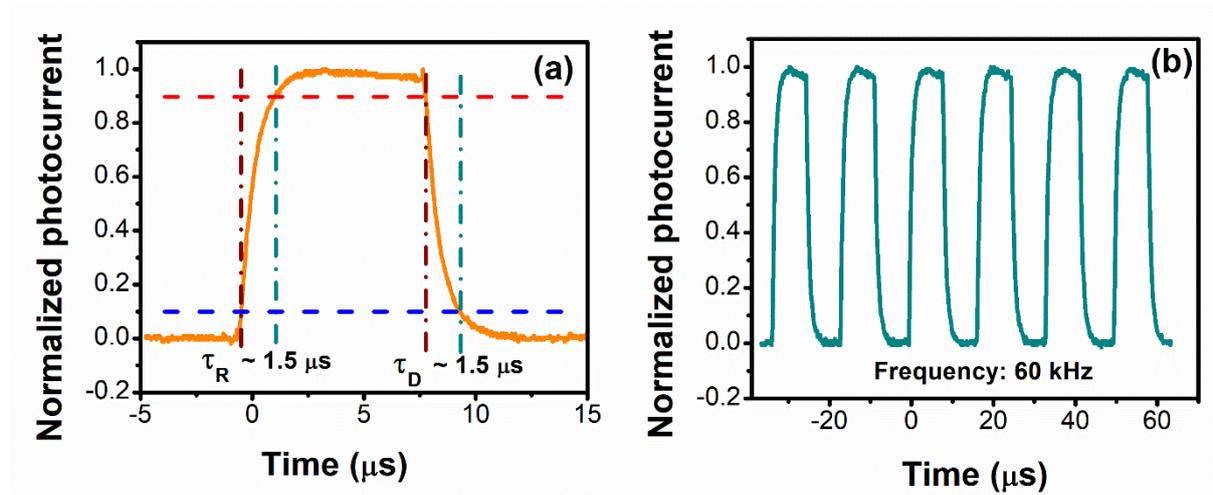

**Figure 5**: (a) Temporal photoresponse of the Au/MgZnO/ITO Schottky junction photodetector under 275 nm UV illumination and (b) a series of normalized photocurrent pulses at 60 kHz showing repeatable behaviour of the device.

As mentioned in the experimental section, the transient response of the Au/MgZnO/ITO Schottky junction photodetectors have been recorded by using a 275 nm UV LED, the output of which has been alternately turned On and OFF at 60 kHz frequency by using square pulse from a function generator. Figure 5 shows the transient photoresponse of the Au/MgZnO/ITO photodetectors, wherein the normalized photocurrent is plotted as a function of time. From figure 5(a) it can be seen that both the rise and decay times of the photocurrent, defined as the time required to rise from 10% to 90% of the peak value and to decay from 90% to 10% of the peak value respectively, are ~ 1.5 μs, implying very fast photoresponse.[10] The fast response observed in this study is due to low junction capacitance of the devices, which has been measured by and LCR meter to be ~ 27 pF at zero bias. Such low junction capacitance has been possible due to the vertical geometry and small cross-sectional area of the devices. Additionally, the relatively high depletion width of ~ 92 nm also contributed in lowering the device capacitance. The relatively large depletion width is plausibly caused by the presence of excess oxygen in the film as revealed by EDAX analysis shown in figure 2(d).[4] Figure 5(b) depicts a series of photoresponse pulses at 60 kHz showing that the photoresponse is highly stable, reliable and reproducible. The fast and stable photoresponse from the Au/Mg$_{0.3}$Zn$_{0.7}$O/ITO Schottky junction photodetectors at a high frequency of 60 kHz imply that such devices can be useful in UV communication.

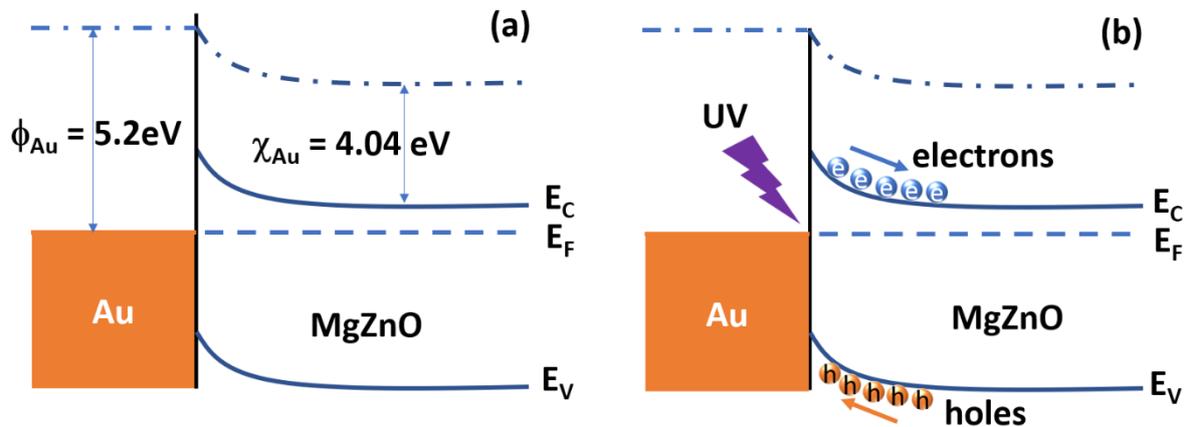

**Figure 6:** Band diagram of the Au/MgZnO Schottky junction under short circuit connection (a) in dark and (b) in UV light.

The mechanism of the observed self-powered visible blind UV photodetection can be explained using the band diagram of Au/MgZnO Schottky junction which is depicted in figure 6 using the Anderson model. Here the work function of Au has been assumed to be ~ 5.2 eV whereas the electron affinity of MgZnO has been assumed from our previous work to be ~ 4.04 eV.[24], [25] Therefor at the Au/MgZnO junction, conduction band of MgZnO will bend upwards creating a depletion region. Under UV illumination with photon energy higher than the bandgap of MgZnO, electron-hole pairs are generated in MgZnO which are separated by the electric field in the depletion region creating the observed photocurrent. Since the MgZnO film, owing to its bandgap, does not absorb visible light, no current is produced under illumination by visible light.

The typical responsivity and response times of MgZnO based Schottky type self-powered UV photodetectors reported in literature are summarized in Table 1. It is clear the Au/MgZnO/ITO Schottky diodes reported in this work have fastest photocurrent rise and decay times, that is also repeatable at 60 kHz frequency. Additionally, the responsivity of the devices at zero bias is comparable to the largest responsivity reported in literature for MgZnO based self-powered UV detectors.

Table 1: Peak responsivity at zero bias and response time of MgZnO based Schottky type self-powered UV photodetectors.

| Photodetector | Rise time | Decay time | Peak responsivity at zero bias |
| --- | --- | --- | --- |
| Au/MgZnO/MgZnO:Al[26] | - | - | 0.0266 mA/W at 340 nm |
| Ag/ZnMgO/ZnO[13] | 0.024 ms | 0.3 ms | 16 mA/W at 275 nm |
| ZnMgO:Al/PEDOT:PSS[15] | 320 ms | 200 ms | 19.1 mA/W at 278 nm |
| Pt/ZnMgO[14] | - | - | 15 mA/W at 220 nm |
| Au/ZnMgO/ZnO:Al[10] | 0.02 ms | 0.3 ms | 55 mA/W at 265 nm |
| Au/MgZnO/Ga:ZnO/In[27] | - | - | 2.55 mA/W at 230 nm |
| Asymmetric Au/MgZnO[28] | - | 92 μs | 2.22 mA/W at 330 nm |
| Gr/MgZnO/4H-SiC[29] | 0.0782 s | 0.162 s | 2.64 mA/W at 255 nm |
| Au/MgZnO/ITO (This work) | 1.5 μs | 1.5 μs | 49 mA/W at 280 nm |

**Conclusion**

In conclusion, MgZnO based self-powered visible blind UV photodetectors have been fabricated in vertical geometry on ITO substrate using Au as the Schottky electrode. By fabricating the devices in crossbar pattern, device cross-section has been made smaller to reduce the junction capacitance for faster operation. The vertical Schottky junctions thus produced have high UV responsivity of ~ 49 mA/W and fast photocurrent rise and decay times of ~ 1.5 μs, that can be useful in UV communication and imaging systems.

**Acknowledgement**

The authors thank Dr. P. Gupta and Dr. S. K. Rai of Accelerator Physics and Synchrotrons Utilization Division of RRCAT, Indore for their help in X-ray diffraction measurements. The authors also thanks Shri A. Chowdhury and Smt. R. Singh of Laser & Functional Materials Division for their help in SEM and EDAX measurements.